\documentclass[aps,preprint,a4paper,showpacs,showkeys,superscriptaddress]{revtex4-1}

\usepackage{latexsym}
\usepackage{amsmath,amssymb}
\usepackage{graphicx}
\usepackage{subfigure}
\usepackage{xcolor}
\usepackage{cancel}

\usepackage{hyperref}     
\hypersetup{colorlinks,%
  citecolor=blue,%
  linkcolor=cyan,%
  pdftex}

\begin{document}


\title{Unruh temperatures in circular and drifted Rindler motions}

\author{Yongwan Gim}%
\email[]{yongwan89@sogang.ac.kr}%
\affiliation{Department of Physics, Sogang University, Seoul, 04107,
  Republic of Korea}%
\affiliation{Research Institute for Basic Science, Sogang University,
  Seoul, 04107, Republic of Korea} %

\author{Hwajin Um}%
\email[]{um16@sogang.ac.kr}%
\affiliation{Department of Physics, Sogang University, Seoul, 04107,
  Republic of Korea}%

\author{Wontae Kim}%
\email[]{wtkim@sogang.ac.kr}%
\affiliation{Department of Physics, Sogang University, Seoul, 04107,
  Republic of Korea}%

\date{\today}

\begin{abstract}
We study the temperatures for the circular and drifted Rindler motions
by employing the Unruh-DeWitt detector method.
In the circular motion, the temperature is increasing along the radius of the circular motion
until it reaches the maximum, and then
it is decreasing and eventually vanishing at the limit to the radius where the proper acceleration is infinite.
In fact, the temperature is proportional to the proper acceleration
quadratically near the origin of the circular motion as compared to the usual Unruh effect
depending on the linear proper acceleration.
On the other hand, in the drifted Rindler motion,
the observer moves with a relative velocity in the direction transverse to the acceleration.
If the detector is moving slowly in
the transverse direction with a finite proper acceleration,
then the temperature
behaves like the usual Unruh temperature,
while it vanishes for the speed of light
in the transverse direction according to the infinite proper acceleration.
Consequently, it turns out that the temperatures behave nonlinearly with respect to the proper acceleration
and the infinite proper acceleration
would not always permit the divergent temperature.

\end{abstract}

%



\maketitle


\section{Introduction}
\label{sec:intro}
A detector in an accelerated frame
measures excitations related to fluctuations of the quantum vacuum,
which results in the remarkable prediction of the Unruh temperature in quantum field theory
\cite{Unruh:1976db}.
The detector is defined to move through a region permeated by a quantum scalar field
and is weakly coupled to the scalar field in its motion,
which is known as Unruh-DeWitt detector
 \cite{DeWitt:1980hx, Birrell:1982ix, Unruh:1983ms}.
Self-correlations of the scalar field on a trajectory cause the detector to observe excitations.
The detector undergoing a uniform linear acceleration $\alpha$ within the Minkowski space perceives
the Unruh temperature as
\begin{equation}\label{eq:TU}
T_{\textrm {\tiny U}} = \frac{\hbar}{k_{\rm B} c}\frac{\alpha}{2\pi}.
\end{equation}
The Unruh temperature of the Rindler observer is equivalent to the temperature measured by a locally fiducial observer near the event horizon of the Schwarzschild black hole in the large mass limit  \cite{Unruh:1976db}.
It is worth noting that
 the experimental verification of the Unruh effect would be difficult,
since the linear acceleration of $2.6\times 10^{22} {\rm cm/s^2}$  is required to produce the temperature of $1{\rm K}$ \cite{Unruh:1998gq}.

Related to the phenomena of the linearly uniform acceleration, one might
wonder what would be the case for a circular motion
having the constant magnitude of a proper acceleration with a constant speed.
 The circular motion is interesting not only from the
theoretical point of view  but also the experimental verification \cite{Bell:1982qr, Bell:1986ir}.
 So,
many efforts have been devoted to studying the rate of the response function
for the circular motion \cite{Letaw:1979wy, Letaw:1980yv, Letaw:1980ik, Takagi:1986kn, Davies:1996ks, Crispino:2007eb,  Gutti:2010nv},
where the Unruh-DeWitt detector is set as a rotating monopole detector for simplifying calculations.
Despite this simplification,
it would be difficult to calculate the rate of response function explicitly.
As a slightly different approach,
the Wightman function was written by a sum over the normal modes,
and then the rate of response function was nicely calculated in Ref. \cite{ Gutti:2010nv}.

On the other hand,
one can consider a similar motion to the circular one such as the
drifted Rindler motion, where the detector is uniformly accelerated
and at the same time moves at a constant speed in the direction perpendicular to the acceleration.
The similarity between the circular motion and the drifted Rindler motion has been discussed in several literatures  \cite{Gerlach:1983iv, Takagi:1986kn, Korsbakken:2004bv},
where the circular motion for the large radius is actually corresponding to the drifted Rindler motion.
For the drifted Rindler detector, it is also difficult to calculate the rate of the response function, 
so that it was treated numerically
and the result is compatible with the Planck spectrum~\cite{Letaw:1980yv}.

It is commonly expected for the Unruh temperature to be proportional to the linear
acceleration of the frame. So, one might wonder what happens for the circular and the drifted Rindler motions.
At first glance, one might expect that the temperatures would still be proportional to the proper acceleration
defined in the circular and drifted Rindler motions.

In this paper, we would like to study the temperatures for the circular motion and
the drift Rindler motion, respectively,
by using  the Unruh-DeWitt detector method.
In the circular motion, the temperature vanishes
not only when the proper acceleration becomes zero but also
when the radius approaches a limit to the radius
where the proper acceleration diverges and the magnitude of the tangential
velocity becomes the speed of light.
On the other hand, in the drifted Rindler motion, if
the detector is moving slowly in
the transverse direction with a finite proper acceleration,
then the temperature
behaves like the usual Unruh temperature.
If the transverse velocity becomes the speed of light, then
the proper acceleration for the drifted Rindler motion goes to infinity and thus
the temperature vanishes very similarly to the case of the circular motion.
Therefore, the temperatures depend on the proper accelerations nonlinearly
in the circular and drifted Rindler motions,
and the infinite proper acceleration
would not always permit that the temperature diverges.

In Sec.~\ref{sec:DM},
we encapsulate the Unruh-DeWitt detector method for the Unruh temperature.
Then, in Sec.~\ref{sec:circular}, we obtain the temperature for the circular motion
by employing the Unruh-DeWitt detector method.
In Sec.~\ref{sec:drifted},
we also calculate the temperature for the drifted Rindler motion.
Finally, conclusion and discussion will be given in Sec.~\ref{sec:con}.
We will set $\hbar= c = k_{\rm B}=1$ for simplicity.

\section{Unruh-DeWitt detector method}
\label{sec:DM}

Let us first recapitulate the Unruh-Dewitt detector method
for the calculations of the Unruh temperatures presented in Refs.~\cite{DeWitt:1980hx, Birrell:1982ix, Unruh:1983ms}.
One should assume a detector moving
through a region permeated by a quantum scalar field $\Phi(x)$ along a trajectory $x^\mu(\tau)$ in the Minkowski spacetime
with a proper time $\tau$.
The Lagrangian of the minimal interaction between the detector and the scalar field
 is written as $L=\kappa \mathcal{D}(\tau) \Phi\left(x(\tau)\right)$
with a small coupling constant $\kappa$, where the detector operator $\mathcal{D}(\tau)$ is defined as
$\mathcal{D}(\tau)=e^{i H_0 \tau} \mathcal{D}_0 e^{-iH_0 \tau}$.

As the detector accelerates,
it will measure the energy transition
from the energy $\mathcal{E}_{\rm i}$ of the ground state to the energy $\mathcal{E}$ of an excited state.
Then, the first order amplitude $\mathcal{A}^{(1)}$ is given by
\begin{align}
\mathcal{A}^{(1)}
& = i \kappa \langle \mathcal{E}| \mathcal{D}_0 |\mathcal{E}_{\rm i} \rangle \int^{\tau}_{\tau_{\rm i}} {\rm d} \tau e^{i \tau (\mathcal{E}-\mathcal{E}_{\rm i})} \langle \psi | \Phi(x) |0 \rangle.
\end{align}
Here, the Minkowski vacuum and the excited state
are denoted as $|0\rangle$ and $|\psi \rangle$ respectively.
So, the transition probability defined as $\mathcal{P} = \int d\mathcal{E} | \mathcal{A}^{(1)}|^2$
is calculated as
\begin{align}
\mathcal{P}
 &= \kappa^2 \int {\rm d}\mathcal{E} |\langle \mathcal{E}| \mathcal{D}_0 |\mathcal{E}_{\rm i} \rangle|^2 \mathcal{R}(\mathcal{E}-\mathcal{E}_{\rm i}),\label{eq:transP}
\end{align}
where the response function $\mathcal{R}$ is
\begin{align}
\mathcal{R}(\Delta \mathcal{E})
&= \int^{\Delta \tau^+}_{\Delta \tau^+_{\rm i}} {\rm d}\Delta \tau^+ \int^{\Delta \tau^-}_{\Delta \tau^-_{\rm i}} {\rm d}\Delta \tau^- e^{-i \Delta \tau^- \Delta \mathcal{E}}  \mathcal{G}^+(\Delta \tau^-)\label{eq:R}
\end{align}
with the energy difference $\Delta \mathcal{E} = \mathcal{E}-\mathcal{E}_{\rm i}$ and the time differences $\Delta \tau^\pm = \tau \pm \tau'$.
Note that the positive frequency
Wightman function $\mathcal{G}^+$ is defined as
\begin{equation}\label{eq:G+}
\mathcal{G}^+(\Delta \tau^-)= \langle 0 | \Phi \left(\tau\right) \Phi\left(\tau'\right) |0 \rangle.
\end{equation}
From the transition probability \eqref{eq:transP},
 we can obtain the transition probability per unit time as
\begin{equation}
\dot{\mathcal{P}} = \kappa^2 \int {\rm d} \mathcal{E} |\langle \mathcal{E}| \mathcal{D}_0 |\mathcal{E}_{\rm i} \rangle|^2 \dot{\mathcal{R}}(\Delta \mathcal{E})
\end{equation}
with the rate of the response function
\begin{align}\label{eq:Rdot}
\dot{\mathcal{R}}(\Delta \mathcal{E}) =  \int^\infty_{-\infty} {\rm d} \Delta \tau^- e^{-i \Delta \tau^- \Delta \mathcal{E}}  \mathcal{G}^+(\Delta \tau^-),
\end{align}
where the integration range of $\Delta \tau^- $ is extended up to $\pm \infty$.

Finally, the Unruh temperature can be read off from the relation of
\begin{equation}
\dot{\mathcal{R}}=\frac{\Delta \mathcal{E}}{2\pi}\frac{1}{(e^{\Delta \mathcal{E}/T}-1)} \label{tem}
\end{equation}
 between the Planck distribution and the rate of the response function
\cite{DeWitt:1980hx, Birrell:1982ix, Unruh:1983ms}.
The rate of the response function is closely related to the temperature of the system
in the sense that it reduces to $\dot{\mathcal{R}}=T/(2\pi)$ for $\Delta \mathcal{E}/T \ll 1$.

\section{Temperature in the circular motion}
\label{sec:circular}
By using the Unruh-DeWitt detector method,
we calculate the temperature measured by the rotating detector.
Let us set the detector to rotate around the $z$-axis with a radius $\rho$
and a constant angular velocity $\Omega$ assumed to be positive finite.
The timelike Killing vector for the rotational motion is given by \cite{Letaw:1980yv, Letaw:1980ik,Padmanabhan:1982, Sriramkumar:1999nw}
\begin{equation}\label{eq:killing}
\xi^\mu = (\gamma,- \gamma \Omega y, \gamma \Omega x,0),
\end{equation}
which is generating the rotational trajectory as
\begin{equation}\label{eq:rot}
x^\mu = (\gamma \tau, \rho \cos(\gamma \Omega \tau),\rho \sin(\gamma \Omega \tau),0),
\end{equation}
where $\gamma=(1-v^2)^{-1/2}$ is the Lorentz factor and
the velocity $v$ is tangent to the circular orbit as defined by $v= \rho \Omega$.

On the other hand, the rotating frame is described by the line element,
\begin{equation}\label{eq:rotating}
{\rm d} s^2 = - (1- \Omega^2 \rho^2) {\rm d}t^2 +2 \Omega \rho^2 {\rm d}\varphi {\rm d} t + {\rm d}\rho^2  + \rho^2 {\rm d} \varphi^2 + {\rm d} z^2.
\end{equation}
Note that
there is a limit to the radius at $ \rho_{\textrm {\tiny H}} = 1/\Omega$ \cite{PhysRev.71.54}.
The proper  acceleration for the rotating motion
is calculated as $a_{\rm cir}=\gamma^2 \rho \Omega^2$,
which is also simplified as $a_{\rm cir} =  \Omega v (1-v^2)^{-1}$.
The proper acceleration vanishes for $v = 0$ corresponding to $\rho = 0$, while the proper acceleration is infinite for $v \rightarrow 1$ corresponding to $\rho \rightarrow \rho_{\textrm {\tiny H}}$.

In the local field theory, the Lagrangian density of the massless free scalar field $\Phi$ is written as
\begin{equation}\label{eq:EOMlocal}
\mathcal{L} = -\frac{1}{2} \Phi(x) (-\square) \Phi(x),
\end{equation}
and the positive frequency Wightman function is obtained as
\begin{align}
\mathcal{G}^+(x,x')
=& \int \frac{{\rm d}^3 k }{(2 \pi)^3 2\omega }  e^{i k_\mu (x^\mu-x'^\mu)}
\label{eq:Gloc} \\
=& \frac{1}{4\pi^2} \frac{1}{\Delta x^\mu \Delta x_\mu}\label{eq:Gloc2}
\end{align}
with $\Delta x^\mu = x^\mu(\tau)-x^\mu(\tau')$.
Plugging the rotating trajectory \eqref{eq:rot} into Eq.~\eqref{eq:Gloc2}, one can get the
explicit form  of the Wightman function
as
\begin{equation}\label{eq:GCM}
\mathcal{G}^+_{\rm cir} (\Delta \tau^-) =\frac{1}{4\pi^2}\frac{1}{ -(\gamma \Delta {\tau^-})^2 +  \frac{4v^4 \gamma^4} {a^2_{\rm cir}} \sin^2\left(\frac{a_{\rm cir}}{2 \gamma} \Delta \tau^-  \right)}.
\end{equation}
In order to get the temperature, one should calculate the rate of response function \eqref{eq:Rdot}
from Eq.~\eqref{eq:GCM}.
However,
it appears to be non-trivial to evaluate the integral ~\eqref{eq:Rdot}
and so the rate of the response function will be obtained in a slightly different fashion
along the line of Refs.~\cite{Davies:1996ks, Crispino:2007eb,  Gutti:2010nv}.

We rewrite the Wightman function by using the cylindrical polar coordinates
in order to get the rate of the response function \eqref{eq:Rdot}.
The momentum is expressed in terms of the cylindrical polar coordinates of  $\theta$ and $k=\sqrt{k_x^2+k_y^2}$ as
\begin{equation}\label{eq:moment}
k^\mu = (\omega, k \cos(\theta), k \sin(\theta), k_z).
\end{equation}
Substituting the circular trajectory \eqref{eq:rot} and the momentum \eqref{eq:moment}
into Eq. \eqref{eq:Gloc},
we can rewrite the positive frequency Wightman function as
\begin{align}
\mathcal{G}^+_{\rm cir}(\Delta \tau^-)
&= \int^\infty_0 \frac{k {\rm d} k}{(2\pi)^2} \int^\infty_{-\infty} \frac{{\rm d} k_z}{2\omega} e^{-i \gamma \omega \Delta \tau^-} J_0 \left(\left|2k \rho \sin\left(\frac{\gamma \Omega}{2}\Delta \tau^-\right)\right|\right),\label{eq:GJ0}
\end{align}
where we used the integral representation of the Bessel function of zeroth order as $J_0(x)=(2\pi)^{-1}\int^\pi_{-\pi} {\rm d} \theta e^{- i x \sin(\theta)}$.
Then, the Wightman function \eqref{eq:GJ0} can be essentially expressed as a sum over the normal modes as \cite{Letaw:1979wy, Letaw:1980ik, Davies:1996ks, Crispino:2007eb, Gutti:2010nv}
\begin{align}
\mathcal{G}^+_{\rm cir}(\Delta \tau^-)
= \sum^{\infty}_{m=-\infty} \int^\infty_0 \frac{k {\rm d}k}{(2\pi)^2} \int^\infty_{-\infty} \frac{{\rm d}k_z}{2\omega} e^{-i \gamma (\omega - m \Omega) \Delta \tau^-} J_m^2 \left(k \rho \right) \label{eq:G+rotF}
\end{align}
by using the relation of
\begin{equation}\label{eq:}
J_0 \left(\left|2k \rho \sin\left(\frac{\gamma \Omega}{2}\Delta \tau^-\right)\right|\right)
=\sum^{\infty}_{m=-\infty}  e^{i \gamma m \Omega \Delta \tau^-} J_m^2 \left(k \rho \right),
\end{equation}
where $J_m(x)$ denotes the Bessel function of order $m$.
Before performing the summation and the integrals in Eq.~\eqref{eq:G+rotF},
we evaluate the integral over the differential proper time $\Delta \tau^-$ of the rate of response function as
\begin{align}
\dot{\mathcal{R}}_{\rm cir}(\Delta \mathcal{E}) =&  \int^\infty_{-\infty} {\rm d} \Delta \tau^- e^{-i \Delta \tau^- \Delta \mathcal{E}}  \mathcal{G}_{\rm cir}^+(\Delta \tau^-) \notag  \\
=& \sum^{\infty}_{m=-\infty} \int^\infty_0 \frac{k {\rm d}k}{2\pi} \int^\infty_{-\infty} \frac{{\rm d} k_z}{2\omega} J_m^2 \left(k \rho \right)  \delta \left(\Delta \mathcal{E} +\gamma (\omega -m\Omega ) \right) \label{eq:Rdotdelta},
\end{align}
 where  $\omega = \sqrt{k^2+k_z^2}$ is always positive.
If we set the energy difference $\Delta \mathcal{E}$ as a positive definite value,
 the delta function in Eq.~\eqref{eq:Rdotdelta} is always zero for the summation over $m$ when $m < \Delta \mathcal{E}/(\gamma \Omega)$.
Thus, the integral over $k_z$ is evaluated as
\begin{align}\label{eq:RdotCM}
\dot{\mathcal{R}}_{\rm cir}(\Delta \mathcal{E})
=& \sum^{\infty}_{m \ge \Delta \mathcal{E}/(\gamma \Omega)} \int^{m\Omega - \Delta \mathcal{E}/\gamma}_0 \frac{k {\rm d}k}{2\pi \gamma} \frac{J_m^2 \left(k \rho \right)}{\sqrt{\left(m\Omega - \Delta \mathcal{E}/\gamma \right)^2 - k^2}}
\end{align}
with $K=\sqrt{(m\Omega - \Delta \mathcal{E}/\gamma)^2-k^2}$.
Note that there exists an upper bound on $k$ as $k \le m\Omega - \Delta \mathcal{E}/\gamma$
to make $K$ a real value.
Finally, the rate of the response function \eqref{eq:RdotCM} is rewritten
by using the hypergeometric function denoted by $_1 F_2 \left[\{\mu\},\{\nu,\lambda\},x \right]$ as \cite{Gutti:2010nv}
\begin{align}
\dot{\mathcal{R}}_{\rm cir}(\Delta \mathcal{E})
= \sum^{\infty}_{m \ge \Delta \mathcal{E}/(\gamma \Omega)} &\frac{\rho^{2m}}{2\pi \gamma}
\left(m\Omega - \frac{\Delta \mathcal{E}}{\gamma} \right)^{2m+1} \notag \\
&\times \frac{_1 F_2\left[\{m+\frac{1}{2}\},\{m+\frac{3}{2},2m+1\},-\rho^2 \left(m\Omega - \Delta \mathcal{E}/\gamma \right)^2\right]}{\Gamma(2m+2)}.\label{eq:RdotFinal}
\end{align}

\begin{figure}[tpb]
  \begin{center}
\subfigure[{~$T_{\rm cir}$ vs $a_{\rm cir}$}]{
  \includegraphics[width=0.48\textwidth]{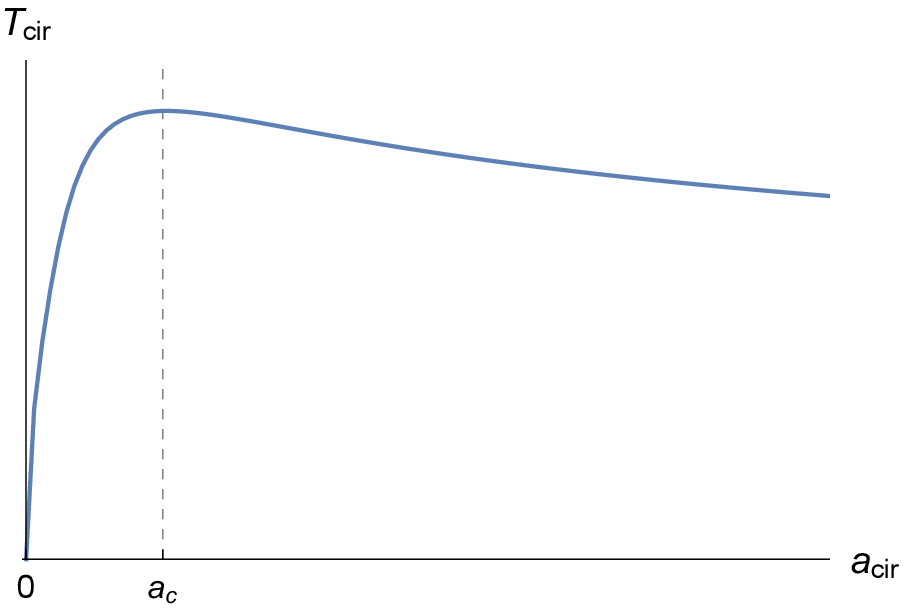}\label{fig:Tvsa}}
\subfigure[{~$T_{\rm cir}$ vs $\rho$}]{
  \includegraphics[width=0.48\textwidth]{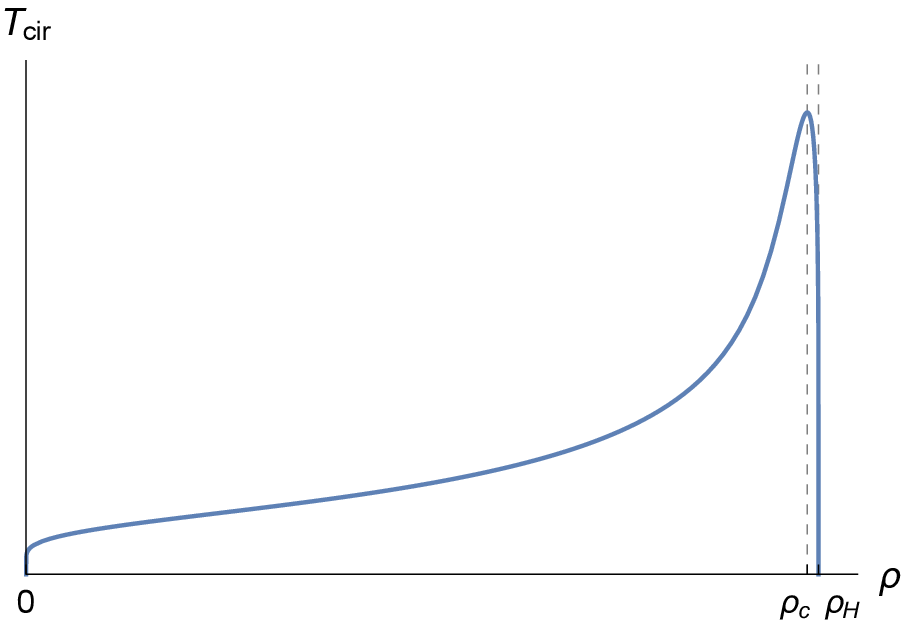}\label{fig:Tvsrho}}
  \end{center}
  \caption{The temperature \eqref{eq:PlanckT} is plotted
  with respect to the proper acceleration $a_{\rm cir}$  in Fig.~(a),
  where 
the temperature vanishes for $a_{\rm cir} \to \infty $ as well as $a_{\rm cir}=0$.
 The behavior of the temperature is also plotted with respect to the radius of the circular
 motion in Fig.~(b) in order to exhibit the whole profile of the temperature easily.
 The temperature vanishes at $\rho=\rho_{\textrm {\tiny H}}$ and $\rho=0$.
 The maximum temperature appears at the critical point $a_{\textrm{\rm c}}$ and $\rho_{\rm c}$.
 Note that the
 parameters are fixed as $\Delta \mathcal{E} =1$ and $\Omega =1$ for convenience.}
  \label{fig:TCM}
\end{figure}


Now, the temperature can be read off from Eqs. \eqref{tem} and \eqref{eq:RdotFinal} as
\begin{align}
T_{\rm cir}= \frac{\Delta \mathcal{E}}{\ln\left(1+ \frac{\Delta \mathcal{E}}{2\pi \dot{\mathcal{R}}_{\rm cir}(\Delta \mathcal{E})} \right)}, \label{eq:PlanckT}
\end{align}
which is plotted in Fig.~\ref{fig:TCM}.
Note that the Unruh temperature \eqref{eq:TU} for
the Rindler motion is proportional to the uniform linear acceleration.
However,
the behavior of the temperature \eqref{eq:PlanckT} shows that
the temperature vanishes at both $a_{\rm cir}=0$ ($\rho =0$) and $a_{\rm cir}\rightarrow\infty$ ($\rho = \rho_{\textrm {\tiny H}}$) as seen from Fig.~\ref{fig:TCM}.
According to this fact, it appears to be natural to exist a maximum temperature in Fig. \ref{fig:Tvsa}
in contrast to the conventional behavior of the acceleration of Unruh temperature for the Rindler motion.

In order to discuss the behavior of the temperature \eqref{eq:PlanckT}
analytically near the center of the circular motion and the limit to the radius,
let us take the leading orders of the response function \eqref{eq:RdotFinal}
and the temperature \eqref{eq:PlanckT} in the IR limit of $\Delta \mathcal{E}\rightarrow 0$.
For the center of the circular motion as $0< \rho \ll \rho_{\rm c} $,
the leading order of the rate of the response function \eqref{eq:RdotFinal} is  written as
\begin{equation}\label{eq:}
\dot{\mathcal{R}}_{\rm cir} = \frac{\Omega }{12\pi} \frac{\rho^2}{\rho_{\rm H}^2},
\end{equation}
where the angular velocity is fixed.
By using Eq.~\eqref{tem},
one can identify the temperature in the IR limit as
\begin{align}
T_{\rm cir}= \frac{v   }{6 }\frac{a_{\rm cir}}{\gamma} \sim a_{\rm cir}^2 \label{eq:PlanckT2},
\end{align}
which is proportional to the proper acceleration quadratically.
So, the temperature for the circular motion is more or less different from
that of the Rindler motion in the IR limit.
On the other hand,
for the limit to the radius as $\rho_{\rm c} \ll \rho < \rho_{\textrm {\tiny H}}$,
the leading order of the rate of the response function \eqref{eq:RdotFinal} in the IR limit is
obtained as 
\begin{equation}\label{eq:Rcir2}
\dot{\mathcal{R}}_{\rm cir} = \frac{\mathcal{B}}{2\pi}\frac{ \Omega }{\gamma},
\end{equation}
where $\mathcal{B}=\sum^{\infty}_{m =1} (\Gamma(2m+2))^{-1} m^{2m+1} ~_1 F_2\left[\{m+\frac{1}{2}\},\{m+\frac{3}{2},2m+1\},-m^2\right] \simeq 17.0161$.
Then the temperature is written as
\begin{align}
T _{\rm cir}
=  \mathcal{B} \frac{a_{\rm cir}}{v \gamma^3} \sim  \frac{1}{\sqrt{a_{\rm cir}}}, \label{eq:PlanckT3}
\end{align}
which is proportional to the inverse square root of the proper acceleration.
This feature is very different from the standard Unruh effect in that the temperature
is no longer proportional to the regular powers of the proper acceleration.
In the next section,
the temperature for the drifted Rindler motion will be explored
how it is different from the Unruh temperature.

\section{Temperature in the drifted Rindler motion}
\label{sec:drifted}

If the radius of the circular motion is very large, the motion looks like the linear accelerated motion with a high speed in the direction perpendicular to the acceleration.
So, we investigate the temperature for the drifted Rindler motion
in order to compare it with the temperature for the circular motion.

Let us consider 
the detector moving with a relative velocity in the direction transverse to the acceleration.
The world line of the drifted Rindler motion is represented by the Rindler coordinates $\tilde{x}^\mu = (\eta,\xi,\tilde{y},\tilde{z})$ \cite{Takagi:1986kn, Russo:2009kw, Kolekar:2012sf},
\begin{equation}\label{eq:DRinR}
\eta(\tau) = \frac{\gamma\tau}{1+\alpha \xi_0}, \quad \xi(\tau) = \xi_0,\quad \tilde{y}(\tau)=v\gamma \tau,\quad \tilde{z}(\tau)=0,
\end{equation}
where $v$ is the velocity of the translational motion along the $y$-direction.
The parameter $\alpha$ means the proper acceleration when the velocity $v$ vanishes.
The constant $\xi_0$ will be fixed as $\xi_0=0$ for simplicity.
From the trajectory \eqref{eq:DRinR},
the drifted Rindler motion in the Minkowski coordinates is described by
\begin{align}
t(\tau) = \frac{1}{\alpha} \sinh\left(\alpha\gamma \tau\right), \quad
 x(\tau) = \frac{1}{\alpha} \cosh\left(\alpha\gamma \tau\right), \quad
  y(\tau) =v\gamma \tau, \quad
   z(\tau) =0,  \label{eq:DRinM}
\end{align}
which recovers the familiar standard Rindler motion for $v=0$.
The line element of the drifted Rindler motion is the same as that of the Rindler one as
\begin{equation}\label{eq:metricR}
{\rm d} s^2 = -(1+\alpha \xi)^2 {\rm d}\eta^2 + {\rm d}\xi^2 + {\rm d} \tilde{y}^2 +{\rm d} \tilde{z}^2,
\end{equation}
where the horizon is located at $\xi_{\textrm {\tiny H}}=-1/\alpha$.
By using the  trajectory \eqref{eq:DRinR} and the metric \eqref{eq:metricR},
the proper acceleration is calculated as
$a_{\textrm {\tiny drift} }= \alpha \gamma^2 $.
It is worth noting that the proper acceleration $a_{\textrm {\tiny drift} }$ diverges
when $v$ approaches the speed of light for a finite non-vanishing linear acceleration of $\alpha$.

Plugging the drifted Rindler trajectory \eqref{eq:DRinM}  into the positive frequency Wightman function~\eqref{eq:Gloc2} in the local field theory,
we can explicitly calculate the Wightman function as \cite{Letaw:1980yv, Takagi:1986kn}
\begin{equation}\label{eq:}
\mathcal{G}^+_{\textrm {\tiny drift}} (\Delta \tau^-) = \frac{1}{4\pi^2} \frac{1}{  v^2 \gamma^2 (\Delta {\tau^-})^2 - \frac{4  }{\alpha^2}\sinh^2\left(\frac{\alpha \gamma}{2}\Delta \tau^- \right)},
\end{equation}
and then the rate of the response function is formally written as \cite{Letaw:1980yv}
\begin{align}
\dot{\mathcal{R}}_{\textrm {\tiny drift}}(\Delta \mathcal{E}) = \frac{1}{4\pi^2} \int^\infty_{-\infty} {\rm d} \Delta \tau^-    \frac{e^{-i \Delta \tau^- \Delta \mathcal{E}}}{  v^2 \gamma^2 (\Delta {\tau^-})^2 - \frac{4}{\alpha^2}\sinh^2\left(\frac{\alpha \gamma}{2}\Delta \tau^- \right)}. \label{eq:RdotDR}
\end{align}
For $v = 0$, the rate of the response function was exactly calculated and
thus the Unruh temperature was identified with $T =\alpha/(2\pi)$ \cite{Unruh:1976db, DeWitt:1980hx, Birrell:1982ix, Unruh:1983ms}.
Unfortunately, it appears to be impossible to evaluate the rate of response function \eqref{eq:RdotDR} exactly for a
finite velocity.

We now express the rate of the response function~\eqref{eq:RdotDR}
for the special limit of
 $\alpha \gg \gamma$ with a finite $v$ or $v \sim 1$ with a finite
 $\alpha$, which eventually results in $a_{\textrm {\tiny drift} } =\alpha \gamma^2\gg 1$.
Then, the second term in the denominator in Eq.~\eqref{eq:RdotDR}
will be dominant, and
the leading term of the rate of response function is approximately calculated as
 \begin{align}
\dot{\mathcal{R}}_{\textrm {\tiny drift}}(\Delta \mathcal{E}) \sim \frac{\alpha}{(2\pi)^2 \gamma}. \label{omega}
\end{align}
From Eq. \eqref{tem}, we obtain the temperature as
\begin{align}
T_{\textrm {\tiny drift}}
& \sim \frac{\Delta \mathcal{E}}{\ln\left(1+   \frac{2\pi \gamma \Delta \mathcal{E}}{\alpha }   \right)}. \label{eq:TDR}
\end{align}
For $\alpha \gg \gamma$ with a finite $v$ and a finite $\Delta E$, the temperature for the drift motion behaves like
\begin{eqnarray}
T_{\textrm {\tiny drift}} \sim \frac{1}{\gamma}T_{\textrm {\tiny U}}, \label{final}
\end{eqnarray}
which reduces to the Unruh temperature, $T_{\textrm {\tiny drift}} = T_{\textrm {\tiny U}}$
for $v=0$, as it should be.
However,
for $v \sim 1$ with a finite $\alpha$,
 the temperature \eqref{eq:TDR} goes to zero, which can be found
in the circular motion
for the limit of $\rho \sim \rho_{\textrm {\tiny H}}$
corresponding to $a_{\rm cir } \gg 1$ as seen from Eq. \eqref{eq:PlanckT3} and Fig. \ref{fig:TCM}.
This feature is compatible with the fact that the motion for the linear accelerated motion with a
high speed in the direction perpendicular to the acceleration looks like the circular motion with a large radius;
precisely, the linear acceleration is related to the angular velocity as $\alpha \sim \Omega$ 
from Eqs.~\eqref{eq:Rcir2} and \eqref{omega}.

\section{Conclusion and discussion}
\label{sec:con}

We calculated the temperatures for the circular and the drifted Rindler motions.
For the circular motion, the temperature is increasing starting from the zero temperature at the
origin of $\rho=0$ corresponding to $a_{\rm cir}=0$, and then it reaches the maximum value of the temperature.
At last, it goes to zero at the limit to the radius of $\rho = \rho_{\textrm {\tiny H}}$ corresponding to $a_{\rm cir}\rightarrow \infty$.
It is interesting to note that the temperature is proportional to the proper acceleration
quadratically near the origin of the circular motion as compared to the usual Unruh effect
depending on the linear proper acceleration.
The asymptotic behavior of the temperature for the limit of $a_{\rm cir}\rightarrow \infty$ is also non-trivial in that
the temperature \eqref{eq:PlanckT3} vanishes in spite of divergent proper acceleration.
On the other hand, we also investigated the temperature in the drifted Rindler motion
for the two special limits satisfying the condition of $a_{\textrm {\tiny drift} } \gg 1$.
First, for $\alpha \gg \gamma$ with a small $v$, the temperature behaves like the Unruh temperature \eqref{eq:TU}
because the spacetime reduces to the Rindler spacetime for $v =0$.
Second, for $v \to 1$ with a finite $\alpha$,
the temperature \eqref{eq:TDR} goes to zero.
Therefore, the infinite proper acceleration
does not always give an infinite temperature
since the Unruh-DeWitt detector could not measure any excitations
for the circular and drifted Rindler motions
as long as the velocity approaches the speed of light.

On the other hand, in the case of the circular motion,
the Killing vector  is null at  the spherical surface of $\rho=\rho_{\textrm {\tiny H}}$,
where it is timelike inside the surface and spacelike beyond the surface \cite{Letaw:1979wy}.
Thus, no object can be at rest relative to the rotating frame beyond the surface,
since there is no object faster than the speed of light.
So, the region beyond  $\rho=\rho_{\textrm {\tiny H}}$  is similar to the region inside the ergosphere surrounding a rotating black hole \cite{Letaw:1980ii}.
It is worth noting that
the particle creation occurs by not only the event horizon but also the ergosphere
 \cite{Starobinsky:1973aij, Unruh:1974bw}.
So, it has been suggested that
the Unruh effects for the detector undergoing the circular motion might be closely related to the ergoregion effect of the rotating black hole \cite{Gerlach:1983iv, Korsbakken:2004bv}
 like the relation between the Unruh effect of the Rindler motion and
 the local thermal effects near the event horizon of the Schwarzschild black hole \cite{Unruh:1976db}.
So, it would be interesting to study the temperature
near the ergosphere of the rotating black hole 
in connection with the results
 presented in this work. 
This issue deserves further attention.

\acknowledgments
We would like to thank Myungseok Eune for exciting discussions.
This work was supported by
the National Research Foundation of Korea(NRF) grant funded by the
Korea government(MSIP) (2017R1A2B2006159).


\bibliographystyle{JHEP}       

\bibliography{references}

\end{document}